\documentclass[%
 reprint,
 amsmath,amssymb,
 aps,
]{revtex4-1}

\usepackage{graphicx}% Include figure files
\usepackage{dcolumn}% Align table columns on decimal point
\usepackage{bm}% bold math
\usepackage{braket}
\usepackage{xcolor}
\renewcommand{\vec}[1]{\boldsymbol{\mathbf{#1}}}
\begin{document}

%\preprint{APS/123-QED}

\title{Spin-polarized fermions with $p$-wave interactions}% Force line breaks with \\
%\thanks{A footnote to the article title}

\author{Furkan \c{C}a\u{g}r\i{} Top}
\thanks{These authors contributed equally to this work}
\affiliation{Research Laboratory of Electronics, MIT-Harvard Center for Ultracold Atoms, and Department of Physics, Massachusetts Institute of Technology, Cambridge, Massachusetts 02139, USA}

 %\altaffiliation[Also at ]{Physics Department, XYZ University.}%Lines break automatically or can be forced with \\
\author{Yair Margalit}%
 %\email{Second.Author@institution.edu}
\thanks{These authors contributed equally to this work}
\affiliation{Research Laboratory of Electronics, MIT-Harvard Center for Ultracold Atoms, and Department of Physics, Massachusetts Institute of Technology, Cambridge, Massachusetts 02139, USA}

\author{Wolfgang Ketterle}
\affiliation{Research Laboratory of Electronics, MIT-Harvard Center for Ultracold Atoms, and Department of Physics, Massachusetts Institute of Technology, Cambridge, Massachusetts 02139, USA}

%\collaboration{CLEO Collaboration}%\noaffiliation

\date{\today}% It is always \today, today,
             %  but any date may be explicitly specified

\begin{abstract}
We study quantum degenerate Fermi gases of  ${^6}$Li atoms at high densities ($10^{15}$ cm$^{-3}$) and observe elastic and inelastic $p$-wave collisions far away from any Feshbach resonance. $P$-wave evaporation reaches temperatures of $T/T_F=0.42$ partially limited by the slow transfer of energy from high to low velocities through $p$-wave collisions.  Via cross-dimensional thermalization, the $p$-wave background scattering volume is determined to be $\lvert V_p \rvert =(39^{+1.3}_{-1.6}a_0)^3$. $P$-wave dipolar relaxation creates a metastable mixture of the lowest and highest hyperfine states.
%\begin{description}
%\item[Usage]
%Secondary publications and information retrieval purposes.
%\item[Structure]
%You may use the \texttt{description} environment to structure your abstract;
%use the optional argument of the \verb+\item+ command to give the category of %each item. 
%\end{description}
\end{abstract}

%\keywords{Suggested keywords}%Use showkeys class option if keyword
                              %display desired
\maketitle

%\section{Highlights}

%\begin{itemize}
%    \item Identical fermions do not have s-wave collision and $p$-wave collisions are suppressed at low temperatures (with $T^2$ scaling\cite{PhysRevLett.82.4208} in the elastic cross-section).
%    \item For the first time, we showed direct evaporative cooling of Fermions without long-range interactions\cite{PhysRevLett.112.010404}. (Ferlaino group evaporated through $p$-wave dipolar elastic collisions) 
%    \item Collisions are the essential part of the evaporative cooling process. We studied elastic and inelastic $p$-wave collisions at high densities.
%    \item We have a record density and a high Fermi energy in a single beam optical dipole trap. 
%    \item We measured the $p$-wave background scattering rate and $p$-wave three body loss rates at low magnetic field. Until now, it was only measured around the $p$-wave Feschbach resonances by looking at three-body losses\cite{MukaiyamaParameter}. 
%    \item We observed dipolar relaxation of $\ket{3/2,+3/2}$ to $\ket{1/2,+1/2}$. It is interesting to see the generation of a steady state impurity population in the lower hyperfine state because the energy release in transfer to lower hyperfine state is $10.95m$K.
%\end{itemize}
\section{Introduction}
Single component Fermi systems are of fundamental interest.  Even for weak interactions, they will show the Kohn-Luttinger instability\,\cite{Kohn1965} at very low temperatures and form Cooper pairs. Since $s$-wave interactions are not allowed in a single-component system, pairing and superfluidity can be unconventional. The $A_1$ superfluid phase of liquid helium-3 is an example of a single-component Cooper pair condensate\,\cite{Kojima2008}, although the spin-polarization at high magnetic fields is at most in the percent range\,\cite{Leduc1990}. Triplet pairing may also occur in the superconductivity of strontium ruthenate\,\cite{Mackenzie2017}.
Due to the Pauli exclusion principle, interactions in single component Fermi gases are very weak at low temperatures. This can be exploited for precision atomic physics measurements, e.g. in optical lattice clocks of strontium and ytterbium\,\cite{RevModPhys.87.637}. However, it is important to understand the weak $p$-wave interactions in such systems which cause shifts of the clock frequency\,\cite{Lemke2011}.

Ultracold gases are metastable systems and decay into molecules via three-body recombination, a process which is quadratic in density. This usually limits the range of densities to around $10^{14}$\,cm$^{-3}$. For many phenomena, density is a scaling parameter - the same phenomenon can be observed at vastly different densities (and correspondingly, at different energy scales or temperatures) \cite{Stringari}.  However, certain phenomenon can disappear at lower densities and set an absolute energy and density scale.  For example, optical properties of Fermi gases change dramatically when the interparticle spacing is comparable to the reduced optical wavelength $\lambdabar$\cite{HELMERSON:90,Busch_1998,Shuve_2009}, in alkali atoms typically 100\,nm, requiring densities around $10^{15}\,$cm$^{-3}$.
In this work, we report on the highest densities ever achieved with ultracold Fermi gases by combining sympathetic cooling with tight compression in an optical potential. Even at densities larger than $10^{15}$\,cm$^{-3}$, we observe lifetimes of more than a second, since the Pauli suppressed three-body recombination coefficient is smaller than for bosonic gases.

In order to maintain or reach quantum degeneracy, elastic collisions are required.  For single component Fermi gases, elastic collisions and thermalization have been observed at temperatures much higher than degeneracy temperatures\,\cite{PhysRevLett.82.4208}, or near $p$-wave Feshbach resonances\,\cite{MukaiyamaParameter} which increase the losses. Here we observe thermalization and evaporative cooling by background $p$-wave collisions in the ultracold regime.
$P$-wave collisions are strongly dependent on the relative momentum $\hbar k$ of the colliding atoms. For $s$-wave collisions, the elastic cross section and the three-body recombination rate are constant near zero temperature, whereas for $p$-wave scattering, they both scale as $T^2$ or $k^4$. In the zero-range limit, the $p$-wave cross section is characterized by the $p$-wave scattering volume $V_p$ as $\sigma_p(k) = 24\pi V^2_p k^4 $ \cite{taylorScattering}.  For sufficiently small $V_p$, the three-body loss rate coefficient $L_3$ has the form $L_3= C (\hbar/m) k^4 V_p^{8/3}$
with a dimensionless scaling constant $C$\,\cite{Suno2003,MukaiyamaScale}.
Evaporative cooling requires a favorable ratio of elastic to inelastic collision. Assuming a Fermi gas with a temperature $T=T_F$ the elastic collision rate $\Gamma_{el}$ is proportional to $V_p^2 k^8$ and three-body loss rate to $V_p^{8/3} k^{10}$ implying a ratio of good-to bad collisions which scales as $1/(C V_p^{2/3} k^2)$. This suggest that a favorable regime is at low density and weak $p$-wave interactions, i.e. far away from any $p$-wave Feshbach resonances. On the other hand, in this regime elastic collisions become very slow, and losses other than three-body decay (e.g vacuum limited trapping time) lead to a maximum ratio of good-to-bad collisions at intermediate densities.  Fig.\,1 shows that for the system studied here, $^6$Li at low magnetic fields there is a favorable window around $T_F= 150\,\mu$K, where the ratio of good to bad collisions peaks around 150, much better than achievable near the $p$-wave Feshbach resonance studied recently\,\cite{MukaiyamaParameter} (Appendix\,\ref{app:good to bad}). 

\begin{figure}
\includegraphics[width=\columnwidth]{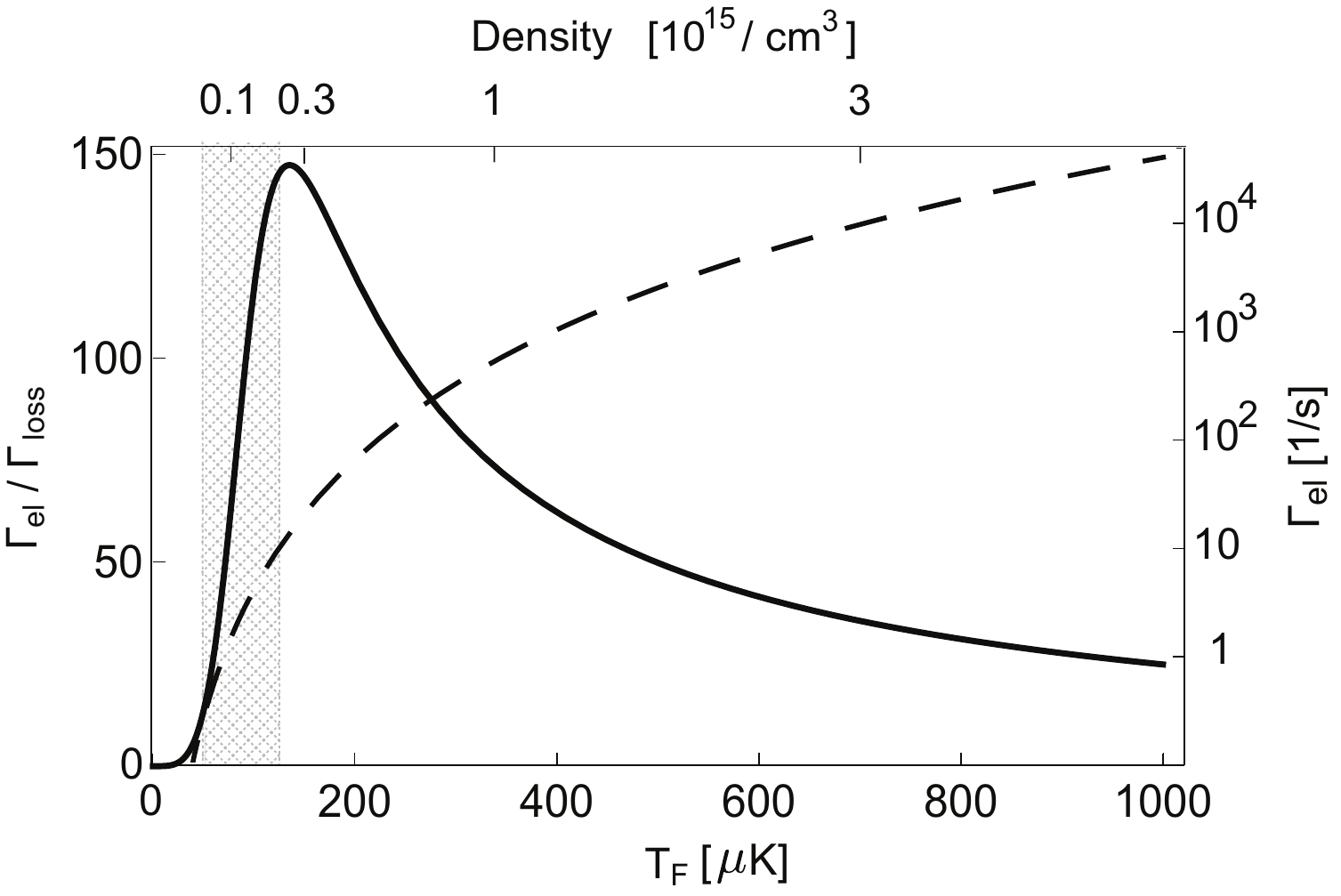}% Here is how to import EPS art
\caption{\label{fig:8} The ratio of good (elastic) to bad (inelastic) collisions, and the elastic collision rate for $^6$Li atoms near zero field, far away from the $p$-wave Feshbach resonance. The solid line shows the ratio for an harmonically trapped cloud with $T=T_F$ assuming a vacuum lifetime of 60 seconds. The dashed line is the elastic collision rate.  In this work, we have explored the shaded region.}
\end{figure}

 After mapping out the collisional properties of $^6$Li, we realized, for the first time, evaporative cooling in the $p$-wave regime. This is different from evaporative cooling with single component dipolar gases (as observed in\,\cite{PhysRevLett.112.010404}), where higher partial waves are not suppressed at low temperatures, and the elastic cross section is constant even at $T=0$, whereas the $p$-wave cross section freezes out proportional to $T^2$.

\section{Sample Preparation} The ultracold lithium clouds are prepared in the following way.  After laser cooling of $^{23}$Na and $^{6}$Li in a double species MOT and  optical pumping of the Li (Na) atoms to the stretched state $\ket{F=3/2,m_F=+3/2}$ ($\ket{2,+2}$), the atoms are captured in a plugged quadrupole magnetic trap \cite{PhysRevLett.75.3969} and sympathetically cooled through forced microwave evaporation of the Na atoms\,\cite{Zoran}. During the last part of the evaporation, a single-beam 1064\,nm optical dipole trap (ODT) with a variable spot size is turned on. The spot size is controlled by a variable-aperture iris shutter, which is initially set to a small open diameter, producing an optical trap with large volume and shallow depth.  This keeps the densities low ($\sim 10^{12}$\,cm$^{-3}$) and avoids inelastic collisions. Finally, the quadrupole field is turned off, and the remaining Na atoms are expelled using a pulse of resonant light. The Li atoms are then transferred to the collisionally stable lowest Zeeman state $\ket{1/2,1/2} \equiv \ket{1}$ using an RF Landau-Zener sweep. Subsequently, the iris is opened to its full aperture in 0.5 sec, thus reducing the ($1/e^2$) ODT spot size radius to approximately $8\,\mu$m and compressing the cloud to densities up to $10^{15}$\,cm$^{-3}$. In order to reduce three-body losses during the iris opening, the 1064 nm laser power is reduced to $30\%$ of its maximum power. The variable spot size ODT is critical to bridge three orders magnitude in density between evaporative cooling and the experiment.

Finally, the ODT power is ramped up to the maximum power of $6.2$\,W. The cloud is held in this tight and deep trap for 30\,ms to ensure thermal equilibrium. A characteristic sample contains $\sim 5.3\times10^6$ Li atoms at $300\,\mu$K temperature and $T/T_F=0.75$, where $T_F$ is the Fermi temperature. The corresponding density is $n=1.3\times 10^{15}$. The trapping frequencies are $(\omega_{rx},\omega_{ry},\omega_z)=(91.5, 102, 2)$ kHz, which are measured by exciting dipole or breathing oscillations. Since $p$-wave interactions are weak, the possibility of impurity populations of Li in other hyperfine states or of Na is a possible concern because they can undergo much faster $s$-wave collisions.  However, these impurities should be rapidly purged from the sample via fast $s$-wave inelastic collisions.  With absorption imaging, we set an upper bound of $0.01\%$ for Na, or for Li in the upper hyperfine state $\ket{F=3/2}$ and in state $\ket{1/2,-1/2}\equiv \ket{2}$.

\section{Three-body loss measurements}  For $^{6}$Li atoms in the lowest hyperfine state, there are no inelastic two body collisions.  Three-body collisions for fermions are suppressed by $k^4$ for three identical fermions, and by $k^2$ for fermions in two states\,\cite{PhysRevA.65.010705}.  So far, $^6$Li three-body recombination has been studied only near a Feshbach resonance which can enhance three-body loss coefficient by six orders of magnitude\,\cite{MukaiyamaScale}, and in a spin mixture with $s$-wave interactions\,\cite{Du2009}.  Due to the high densities achieved here, we are sensitive to background $p$-wave losses (i.e. at a field of 1\,G, far away from the 1-1 Feshbach resonance located at 159\,G).

\begin{figure}
\includegraphics[width=\columnwidth]{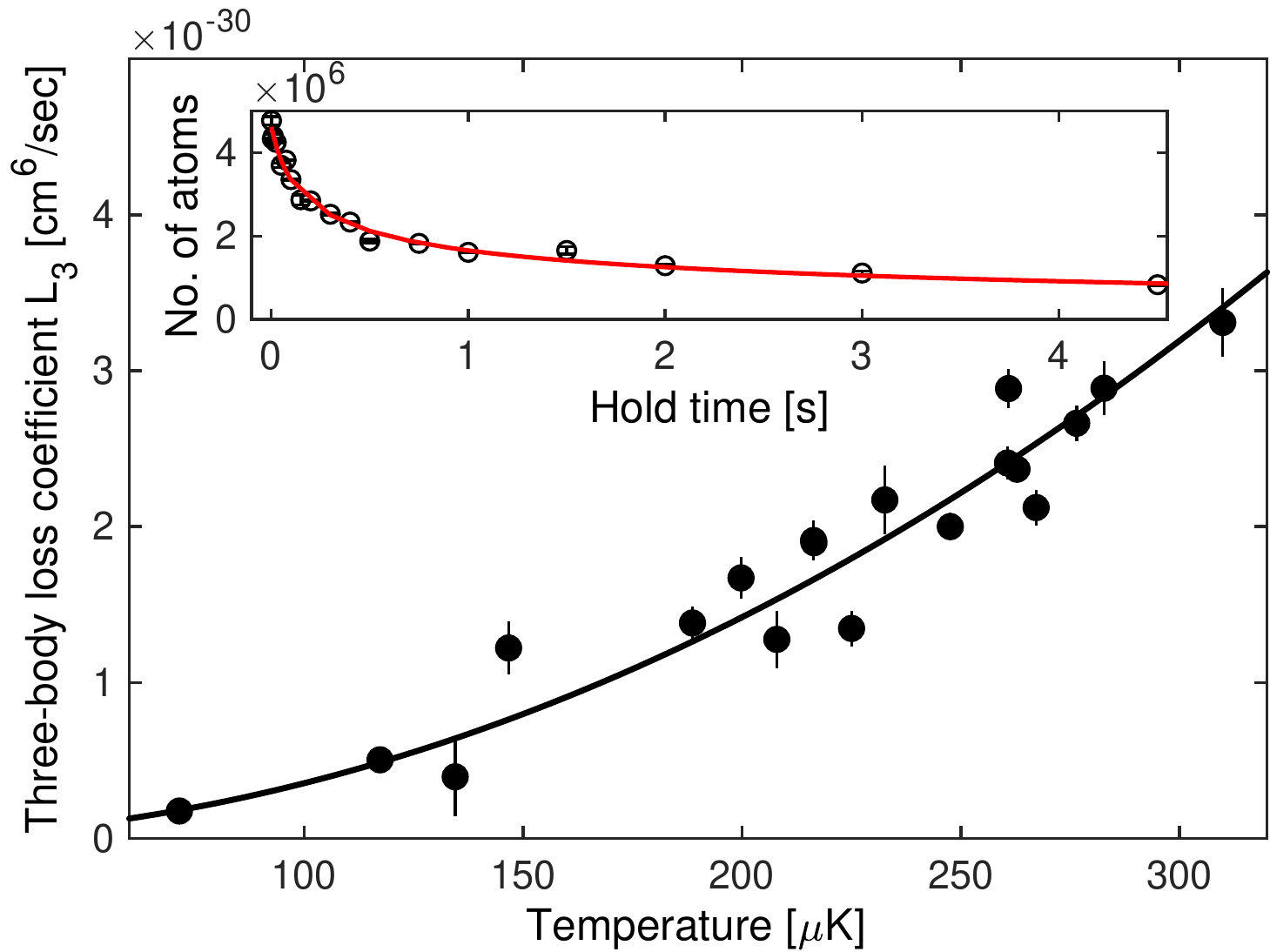}
\caption{\label{fig:3} Three-body loss for spin polarized fermions. Different temperatures are realized by reducing the trap laser power.   The three-body loss coefficient $L_3$ scales quadratically with temperature. The black line is a parabolic fit through the origin, resulting in $L_3 = (3.55\pm0.22)\times10^{-23}\times  T^2$. Error bars show the statistical one standard deviation uncertainties of the fit parameter $L_3$ used to describe the atom number decay.  The inset shows the decay curve at $T=310\,\mu$K, and its fit to Eq.\,\ref{eq:loss}.}
\end{figure}

We observe three-body decay by monitoring the decrease in the number of trapped atoms  (Fig.\,\ref{fig:3}). The rate of change of the number of trapped atoms $N$ is given by
\begin{equation}\label{eq:loss}
\dot{N} = L_1 N -L_3\braket{n^2} N.
\end{equation}
where the density independent losses, parametrized by $L_1$ are almost negligible with $1/L_1 = 31.1\pm1.3$\,sec.  The actual analysis accounts for changes in temperature and cloud size as a function of hold time, and the anharmonic trapping potential (Appendix\,\ref{app:three body}). The results in  Fig.\,\ref{fig:3} confirm the quadratic scaling of $L_3$ with temperature according to a Wigner threshold law\,\cite{PhysRevA.65.010705}, as already observed near a $p$-wave Feshbach resonance \cite{MukaiyamaScale}. At 100 $\mu$K temperature, we obtain $L_3= (3.55\pm0.22)\times10^{-31}\,$cm$^6$/sec, which is among the smallest three-body rate coefficients observed for ultracold atoms, illustrating the high stability of spin polarized Fermi gases. Thermal gases of sodium and rubidium have rate coefficients between $10^{-29}$ and $10^{-28}\,$cm$^6$/sec \,\cite{DSKurn,F2Na3B, Soding1999}.

Using the data in Fig.\,\ref{fig:3}, we obtain a dimensionless scaling constant value of $C=(3.0\pm0.6)\times10^{4}$, compared to the value of $C=2\times10^6$ reported in Ref.\,\cite{MukaiyamaScale} near the Feshbach resonance. While bosons show a universal character of three-body losses (i.e. the loss coefficient is independent of the details of the interatomic potential\,\cite{Nielsen1999}), the different $C$ coefficient values demonstrate for the first time the lack of universal character in recombination of three ultracold fermions, as has been hypothesized in Ref.\,\cite{Suno2003}.

\section{Thermalization measurements} Elastic collision rates are observed by creating a non-equilibrium state and monitoring the collisional relaxation back to equilibrium \,\cite{MonroeCD, PhysRevLett.82.4208}. In order to avoid systematics due to three-body losses, heating, and trap anharmonicities, the atom number is lowered to three different values around 10\% of its maximum value. An anisotropic temperature (with up to $35\%$ temperature difference between the radial and axial directions) is created by spilling atoms out of the trap in the axial direction by by decreasing the trap laser power and adding a magnetic field gradient along the axial direction.  The magnetic gradient is then turned off and the ODT power is non-adiabatically ramped back up to the full power. Thermal relaxation by $p$-wave collisions is observed by monitoring the time evolution of the temperature difference between axial and radial directions.

\begin{figure}
\includegraphics[width=\columnwidth]{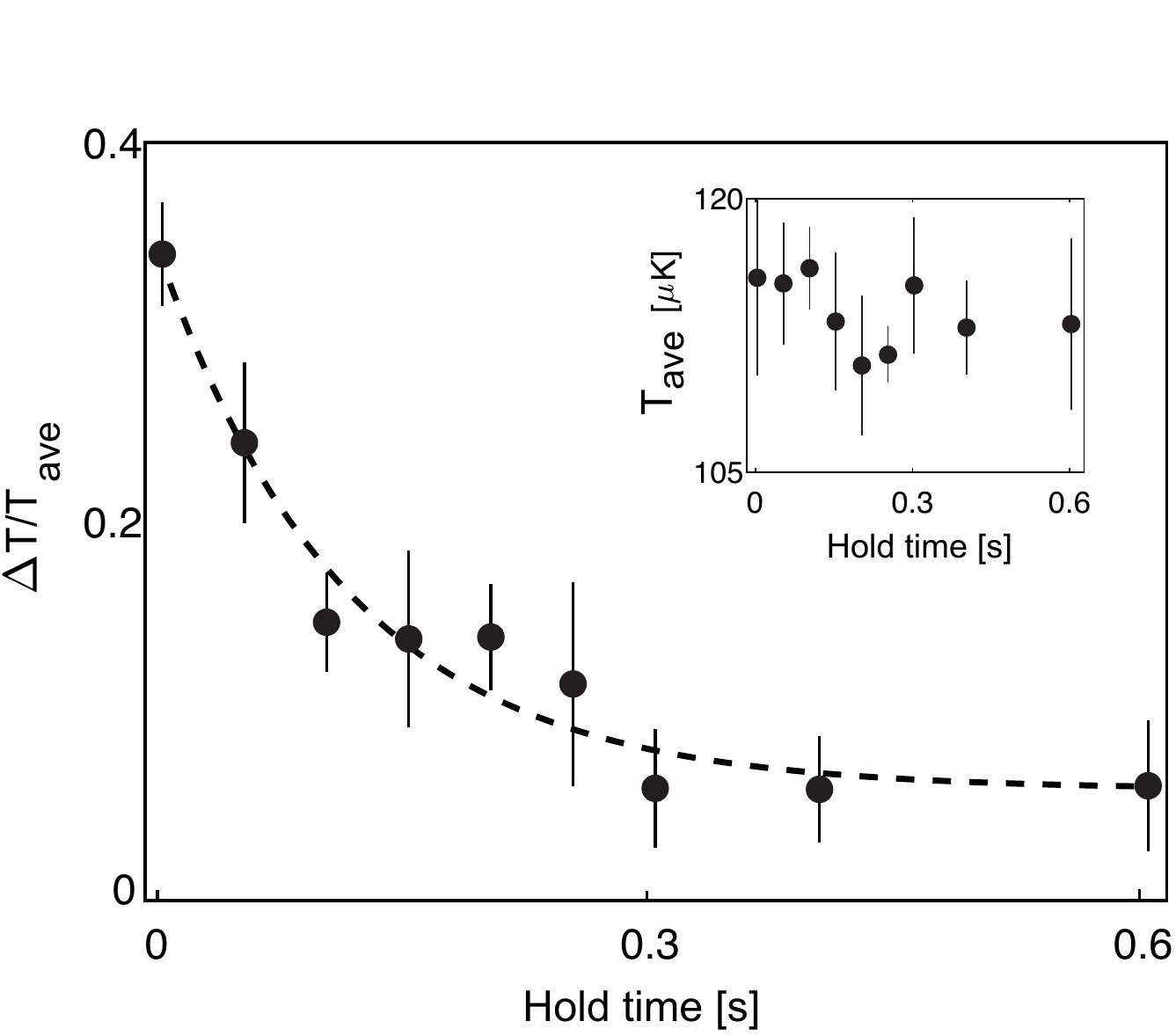}
\caption{\label{fig:thermalization} Observation of cross-dimensional thermalization. Normalized temperature difference $\Delta T/T_\text{ave}$ between the radial and axial directions as a function of hold time after creating a sample with anisotropic energy distribution.  The dashed line is an exponential fit $\Delta T/T_\text{ave} = A \exp(-\Gamma_{th} t)+c$.  
%We note that the measured temperature difference does not decay to zero because the trap turn-off is not completely sudden with respect to the radial trap frequencies, which creates 
The temperature difference has a small (5\%) offset in all measurements the reason for which we have not tracked down (possibly due to anisotropic heating/cooling or non-sudden switch-off of potentials to initiate ballistic expansion).  Inset shows the average temperature, demonstrating that there is no heating during the measurement. Error bars represent the  standard deviation of the data averaged over 3 measurements.}
\end{figure}

%The $p$-wave thermalization rate is given by\,\cite{MukaiyamaParameter} $\Gamma_{th} = \frac{2}{\alpha}\braket{n^2 \sigma(E) v_r E}/k_B T$, where $n$, $\sigma(E)$, and $v_r$ are the atomic density, the $p$-wave elastic scattering cross section for the relative scattering energy $E$ and the relative velocity, respectively, and where $\alpha=4.1$ is the average number of $p$-wave scatterings per atom required for thermalization\,\cite{PhysRevLett.82.4208}. The bracket denotes thermal averaging. The relative scattering energy is given by $E=\frac12\mu v_r^2$, where $\mu$ is the atom's reduced mass. At low energies, and away from a Feshbach resonance, the p-wave scattering cross-section for identical fermions is given by $\sigma_p(k) = 24\pi V^2_p k^4 $, where $k$ is the wave number associated with the atom. Using the harmonic approximation to calculate the density, the thermalization rate becomes:

A typical thermalization measurement is shown in Fig.\,\ref{fig:thermalization}. The thermalization rate $\Gamma_{th}$ is obtained from an exponential fit to the decay curve of the temperature difference.  The thermalization rate is proportional to the elastic collision rate ($\Gamma_{th}=\Gamma_{el}/\alpha$) where $\alpha$ is the average number of elastic collisions necessary for the cross-dimensional thermalization. An analytical calculation of the energy exchange in a Boltzmann gas gives $\alpha = 64/15$ for $p$-wave collisions\,\cite{ThermElastic} which agrees with a value $\alpha=4.1$ obtained from Monte-Carlo simulations\,\cite{PhysRevLett.82.4208}.  Using the thermally averaged $p$-wave elastic collision rate, we obtain an expression for the thermalization rate (see Appendix\,\ref{app:therm calc} for the full derivation)
\begin{equation}\label{eq:GammaTh}
    \Gamma_{\text{th}} = \frac{1152}{\alpha}\sqrt{2\pi}\frac{V^2_p \mu^{3/2}}{\hbar^4}n_{\text{mean}}(k_B T)^{5/2},
\end{equation}
where $\mu$ is the reduced mass of two $^6$Li atoms, and $n_{\text{mean}}$ the average atomic density. Here we assume that contributions from an effective range expansion are negligible, which should be well fulfilled in the low $k$ limit\,\cite{Gautam2010}, specifically when $k |V_p^{1/3}|\ll1$\,\cite{Idziaszek2009}, which is valid in our case.

\begin{figure}
\includegraphics[width=\columnwidth]{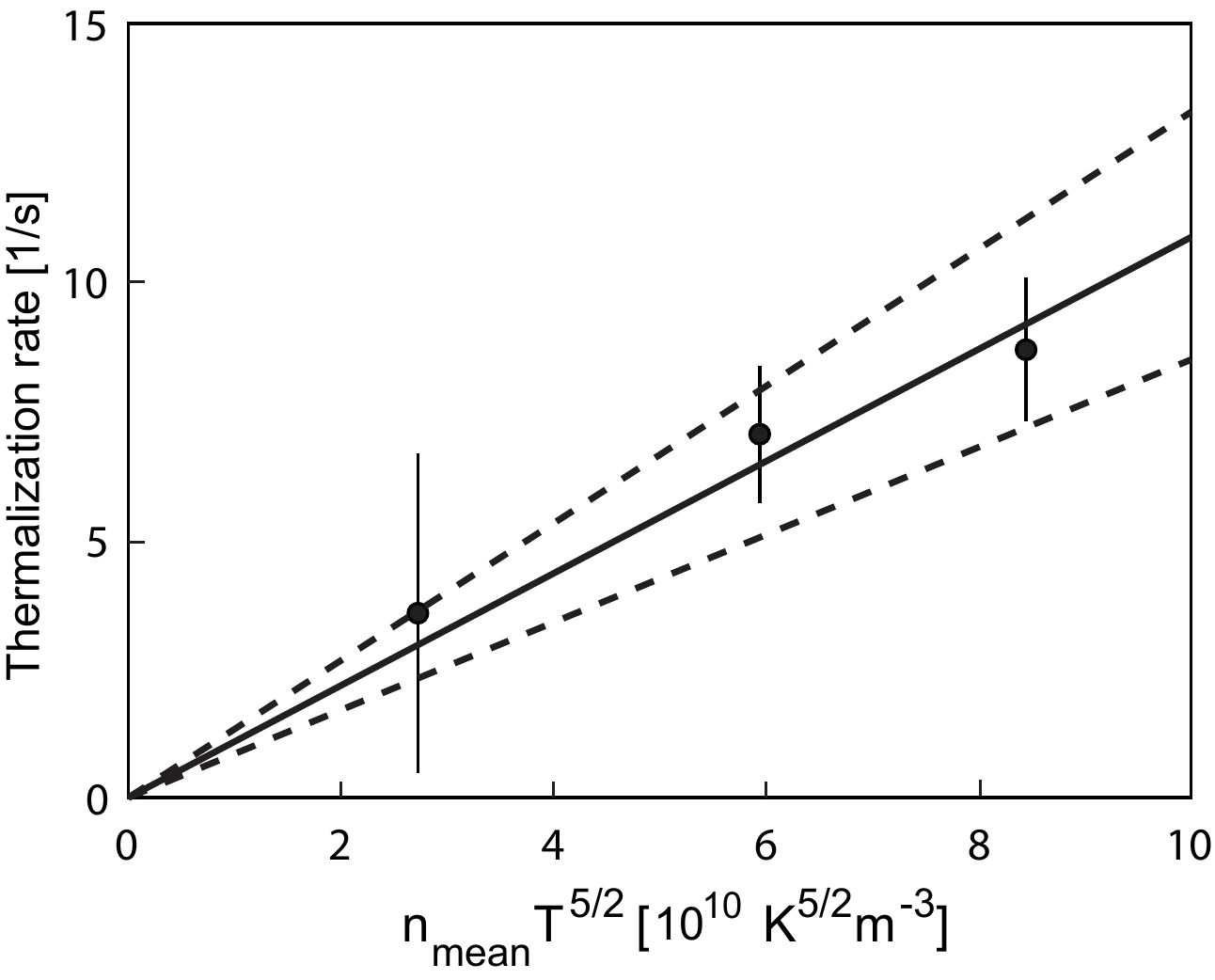}
\caption{\label{fig:scattering length} Determination of the background $p$-wave scattering length. Thermalization rates $\Gamma_{\text{th}}$ were measured for a range of atom numbers and temperatures and are shown as a function of the mean density times $T^{5/2}$, which is the scaling with density and temperature given by Eq.\,\ref{eq:GammaTh}. A linear fit weighing the data with inverse of variances (vertical error bars) gives the $p$-wave background scattering length $\lvert V_{p} \rvert = (39^{+1.3}_{-1.6}\;a_0)^3$. The dashed lines represent the 95\% confidence interval of the fit.}
\end{figure}

Figure\,\ref{fig:scattering length} shows the measured thermalization rates as a function of $n_{\text{mean}}T^{5/2}$. Using a linear fit to the data, the  background $p$-wave scattering volume of state $\ket{1}$ is determined to be $\lvert V_{p} \rvert =(39^{+1.3}_{-1.6}\;a_0)^3$, where $a_0$ is the Bohr radius. The measured value is in reasonable agreement with theoretically calculated values of  $(-35.3a_0)^3$\,\cite{UedaEffRange} and $(-36a_0)^3$\,\cite{Gautam2010} and is the first experimental measurement of a background scattering volume\,\cite{footnote}. This measurement was done at a magnetic field of 1\,G, far away from the 1-1 Feshbach resonance located at 159G.  However, there may still be a small contribution from the wings of the resonance. Assuming a resonance width of $\Delta B=40\,$G\,\cite{AustenThesis}, the scattering volume at $B=1$\,G is decreased by $\Delta B/(B-B_{\text{res}})$ or 25\% from the background value, yielding a corrected background scattering volume of $(42 a_0)^3$.
%we obtain $V(B)/V_p = 1 + \Delta B/(B-B_{res}) = 0.75$, yielding $|V_p^{1/3}|=42 a_0$.  %Since the number of atoms is lowered to around 10\% of the full value, anharmonic corrections to the density do not play a role here.

%CALCULATE AND DISCUSS SCALING AND THE C PARAMETER, COMPARE TO MUKAIYAMA.  DISCUSS THAT OUR DATA IS TAKING IN THE SMALL K LIMIT WHERE THE SCATTERING VOLUME IS THE ONLY TERM NECESSARY

\section{Evaporative Cooling} For a range of temperatures and densities, the measured elastic collision rates are much higher than the inelastic loss rates, allowing for evaporative cooling, using the following procedure.  After sample preparation, the optical trap is ramped down to a fifth of its original value, where the ratio of good to bad collisions is around 70, and the ratio of the trap depth to temperature is $\eta=7$. The trap depth is reduced by a magnetic force applied in the axial direction which is ramped up to various values during 1.5 second.  This ``tilt evaporation'' maintains the confinement and is superior compared to simply reducing the laser beam power \cite{PhysRevA.78.011604}. The square root time dependence of the ramp follows the decreasing elastic collision rate.  After turn-off of the magnetic field gradient, the optical trap is restored to its full value, and a time-of-flight absorption image measures the velocity distribution.  By performing the analysis always with atoms released from the original trap, we avoid systematic effects in comparing different trapping potentials, e.g. anharmonicities.

% The fugacity (directly related to $T/T_F$) and temperature of the cloud are determined from the shape and size of Fermi-Dirac fits (represented by a polylogarithmic function\,\cite{VarennaNotes}) to the time-of-flight distribution of the released atoms.

\begin{figure}[!]
\includegraphics[width=\columnwidth]{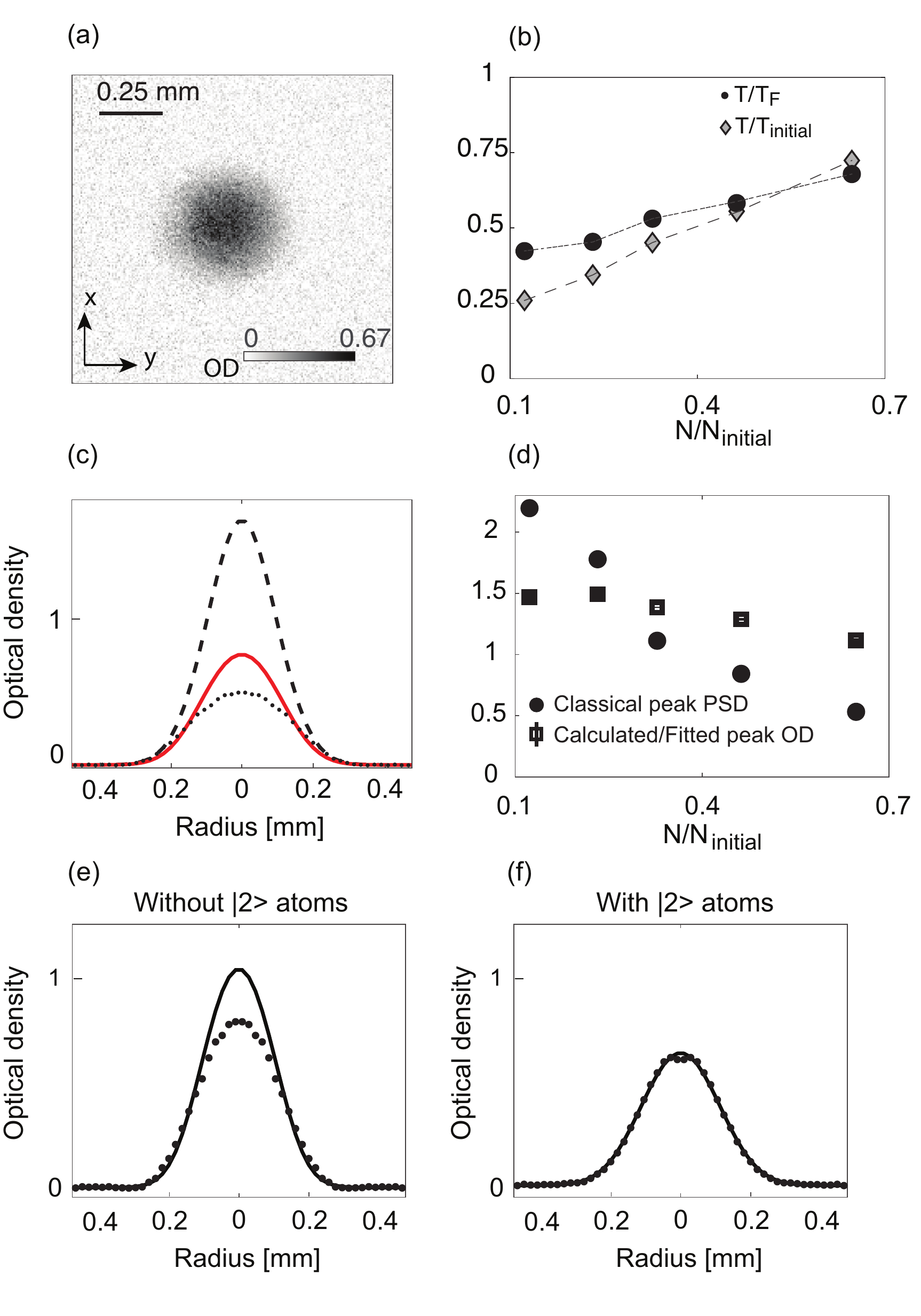}
\caption{\label{fig:5} Evaporative cooling using $p$-wave collisions. (a) Time-of-flight absorption image (average of 4 shots) of degenerate atoms at $T/T_F=0.42$. Temperature is determined from a Gaussian fit to the wings of the cloud, and Fermi energy is calculated using measured number of atoms and trapping frequencies\,\cite{VarennaNotes}. (b) Evolution of $T/T_F$ for different final trap depths and therefore different final numbers of atoms. (c) Gaussian fit (dashed line) to the outer wings (radius $>1.7\sigma$) of the atomic cloud in panel (a), shown as an azimuthally-averaged radial profile. The solid red line shows the equilibrium Fermi-Dirac distribution for the measured temperature and atom number. Its deviation from the data (black dots) demonstrates the insufficient thermalization of the low velocity atoms. (d) Classical phase-space density evolution during evaporation, and the ratio of the calculated and observed peak optical density (squares), demonstrating an increase in the depletion of low-velocity atoms.  (e), (f) Comparison of the azimuthally averaged radial cloud profiles, before and after introducing a second spin component of state $\ket{1/2,-1/2}\equiv \ket{2}$ which undergoes $s$-wave collisions with state $\ket{1}$. Black line shows the Fermi-Dirac profile set by the atom number, temperature and trapping frequencies. The cloud in (e) is not in thermal equilibrium, in contrast to (f).   Note that the number of atoms in (f) is lower, probably due to enhanced three-body loss.}
\end{figure}

We monitor the results of evaporation by determining the final temperature from a Gaussian fit to the wings of the time-of-flight distribution, and the Fermi temperature from the measured number of atoms $N$ and trap frequencies (according to $k_B T_F=\hbar(\omega_{rx}\omega_{ry}\omega_z 6N)^{1/3}$). Fig.\,\ref{fig:5} b and d show a decrease in temperature and $T/T_F$ and an increase in the peak classical phase space density $\rho_{\text{class}}=\frac{1}{6}(\frac{T_F}{T})^3$ as the final trap depth is lowered (by increasing the magnetic force).  The calculated ratio of good to bad collisions decreases from 70 to 23.  The increase in degeneracy is rather modest, since a ratio of good to bad collisions of around 50 will allow evaporative cooling only with modest efficiency.  For this ratio, the $\gamma$ parameter (increase in the logarithm of the phase space density over logarithm of atom loss) is at best around 1 \,\cite{Ketterle1996}.

The clouds after evaporation are not fully equilibrated:  The Fermi-Dirac profile calculated with the fitted values for $T$ and $T_F$ do not match well with the observed profiles. The deeper the evaporation progresses, the larger is the ratio of calculated and observed peak optical density.  We ascribe the lower population for  low velocities to the strong velocity dependence of $p$-wave collisions (see Appendix\,\ref{app:s and p comp}).  Equilibrium in the cloud is quickly established when we use a diabaic Landau-Zener sweep to admix $10\%$ spin impurities (atoms in state $\ket{2}$) right after the evaporation ramp which undergo $s$-wave collisions with the atoms in state $\ket{1}$. For this mixture, we always observe fully equilibrated Fermi-Dirac distributions  (Fig.\,\ref{fig:5}(f)).
% for which $T/T_F$ derived from the atom number and  temperature of the wings  agrees with the $T/T_F$ value obtained from a fit to the shape of the atoms' velocity distribution.

We note that we have observed non-equilibrium distributions also after cross-sectional thermalization (as measured in Fig.\,\ref{fig:thermalization}) showing that full thermalization takes longer than the so-called thermalization time for distributing energy isotropically, an effect which has not been pointed out before.  We conclude that $p$-wave collisions are less efficient for evaporative cooling than $s$-wave collisions, and require a considerable higher ratio of good to bad collisions.

% The cosine square angular distribution requires 1.5 \textcolor{red}{WK:  Shouldn't it be 1.5?  =4.1/2.7} as many collisions to equilibrate kinetic energy between orthogonal directions \cite{ThermElastic}. In addition, the $v^4$ dependence of the $p$-wave cross section implies that four times as many collisions are needed to transfer population to the lowest velocities (see supplement), an effect which has not been pointed out before.  Therefore, efficient $p$-wave evaporation requires a substantially higher ratio of good-to-bad collisions than $s$-wave evaporation.

\section{Dipolar relaxation of state $\ket{6}$}  In the magnetic trap, we cool lithium atoms in state $\ket{3/2,3/2}\equiv\ket{6}$. By omitting the Landau-Zener sweep in the sample preparation, we can study collisional properties of state $\ket{6}$ clouds.  Initial measurements on thermalization rates found the surprising result that thermalization in state $\ket{6}$ is much faster than in state $\ket{1}$.  It turned out that this is due to an admixture of state $\ket{1}$ atoms in a quasi-equilibrium concentration.  This is caused by an amazing interplay of $p$-wave dipolar relaxation of state $\ket{6}$ atoms, ultrafast ($\mu$s) $s$-wave spin relaxation between state $\ket{6}$ and state $\ket{3/2,1/2}\equiv\ket{5}$ atoms in which state $\ket{5}$ decays to state $\ket{1}$, fast collisional relaxation by $s$-wave elastic collisions between states $\ket{1}$ and $\ket{6}$ and loss of state $\ket{1}$ atoms by three-body recombination.

\begin{figure}[!]
\includegraphics[width=\columnwidth]{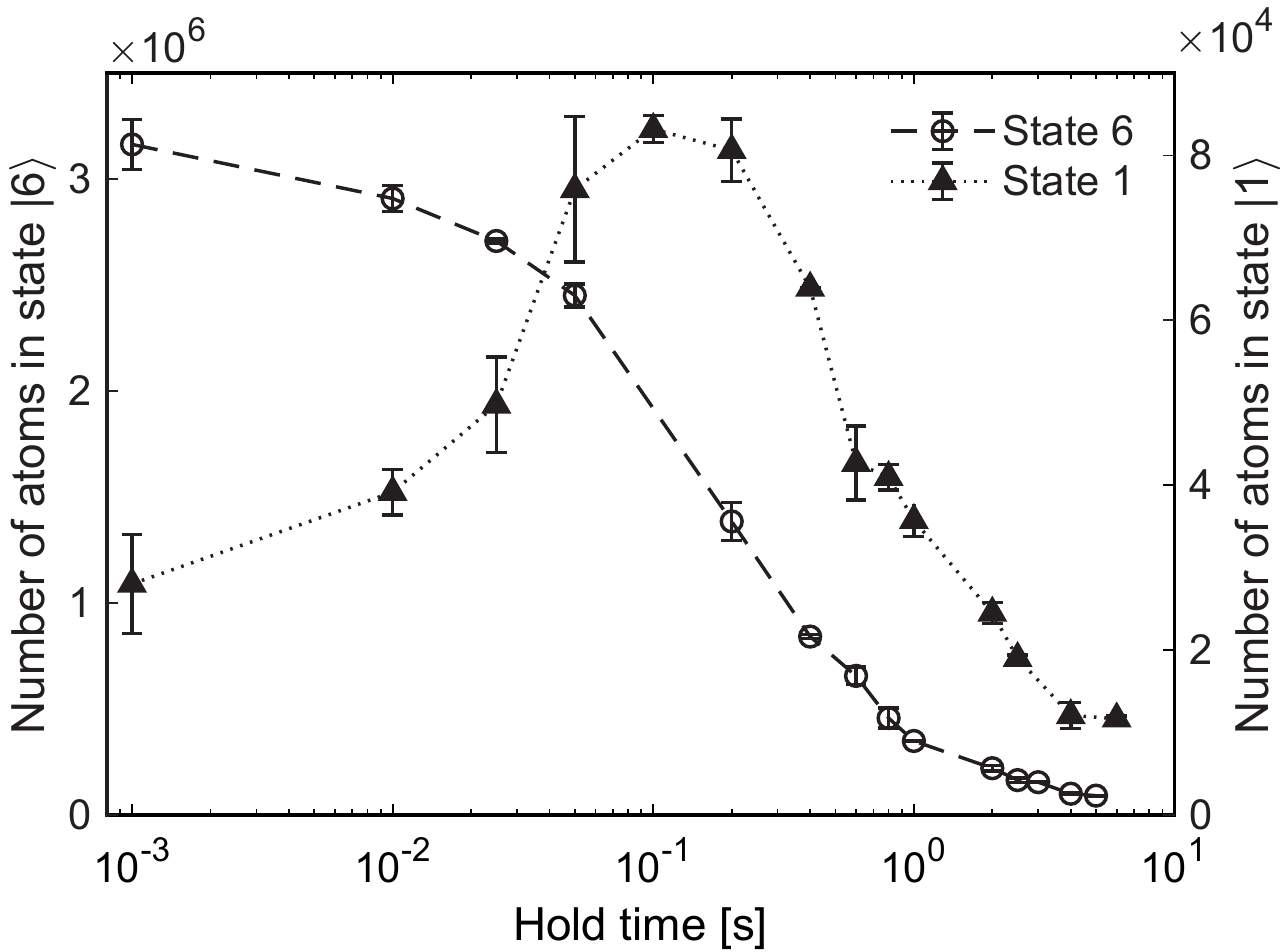}
\caption{\label{fig:dipolar} Dipolar relaxation of state $\ket{6}$, leading to population of state $\ket{1}$. Population of both states, as a function of hold time in the trap at full power. Lines are a guide to the eye. The initial number of state $\ket{1}$ atoms are generated while the optical trap was ramped up to full power.}
\end{figure}

Via absorption imaging, we find that a pure state $\ket{6}$ cloud creates a growing admixture of state $\ket{1}$ atoms (Fig.\,\ref{fig:dipolar}).  Although our detection scheme is sensitive to both states $\ket{1}$ and $\ket{2}$ of the lower $F=1/2$ hyperfine level, we can exclude the presence of state $\ket{2}$ since it undergoes rapid spin relaxation with state $\ket{6}$ (with a rate coefficient of $10^{-8}\,$cm$^3$/s at 1\,G field\,\cite{VanAbeelen1997}).

The dominant decay mechanism for state $\ket{6}$ atoms is three-body loss, similar to the loss observed in state $\ket{1}$. Indeed, from the initial slope of the decay, we obtain a loss rate of $\dot N/N=-6.3$/s, slightly larger than observed for state $\ket{1}$ under the same conditions.  A second, weaker loss mechanism is $p$-wave dipolar relaxation which transfers spin angular momentum to orbital angular momentum \,\cite{Hensler2003}. The rate coefficient for producing state $\ket{5}$ atoms is calculated to be $L_2^{\text{dip}} = 3.2\times10^{-16}\,$cm$^3$/s\,\cite{Hensler2003}. A generated state $\ket{5}$ atom undergoes rapid spin relaxation with state $\ket{6}$ atoms to state $\ket{1}$ with a predicted rate coefficient of $2\times10^{-9}\,$cm$^3$/s\,\cite{SpinExchange}.  At our densities, this spin flip takes place in less than 1 $\mu$s, one of the fastest collisional processes ever observed with ultracold atoms. This spin flip releases the ground state hyperfine splitting energy of 11\,mK (3 times the trap depth). Our detection of cold and trapped state $\ket{1}$ atoms implies that the cloud is collisionally dense for collisions between states $\ket{1}$ and $\ket{6}$.  A collisional density of one along the radial direction requires an $s$-wave scattering length of around $200a_0$.  We don't know of any predictions for 6-1 scattering length, but the 6-2 scattering length is predicted to be large (around $-1700 a_0$\,\cite{VanAbeelen1997}) due to the large triplet scattering length for $^6$Li of $-2160a_0$\,\cite{SpinExchange}. Note that all states should decay rapidly in spin relaxation collisions with state $\ket{6}$, except for state $\ket{1}$ for which spin relaxation is not possible since there are no lower lying states with the same quantum number of total spin along the z direction.

The initial growth rate in the number of state $\ket{1}$ atoms of $\dot{N}_1 = 1.06\times 10^6$ atoms/s is in reasonable agreement with the predicted production of state $\ket{5}$ atoms by dipolar relaxation, and implies a capture efficiency close to unity.  After 1\,s, the relative population in state $\ket{1}$ reaches a quasi-equilibrium, indicating a loss mechanism for state $\ket{1}$ with a rate of around 10/s.  State $\ket{1}$ can disappear in three-body recombination with possible combinations of states 6,6,1 or 6,1,1 or  1,1,1.  Since some of the rate coefficients are not known, we can't model the dynamics.  The observation of a quasi-constant fraction of state $\ket{1}$ atoms (in contrast to a constant number) suggests that the ratio of production and loss rate is proportional to the density of state $\ket{6}$ atoms.

\section{Discussion}
In this paper we have explored a spin-polarized Fermi gas at unprecedentedly high densities and observed various collisional interactions.  Even at the high Fermi temperatures, the elastic collision cross section of $9\times10^{-16}\,$cm$^2$ is four orders of magnitude smaller than typical $s$-wave cross sections of $7\times10^{-12}\,$cm$^2$.  Nevertheless, due to the high density, these Fermi gases equilibrate on time scales of 100 ms, but otherwise are weakly interacting: the $p$-wave mean field energy at zero temperature and $10^{15}\,$cm$^{-3}$ density is 0.2\% of the Fermi energy\,\cite{Roth2001}.

A critical parameter is the ratio of inelastic to elastic collisions which is high enough to allow evaporative cooling. An optimized trap and improved vacuum lifetime should result in even deeper degeneracy.  %\emph{(Note: I am not comfortable quoting a number here. Let me explain. i) At 3.5V trap, we have $T_F=400\mu$K. We evaporate at 0.6V. Adiabatic decompression to 0.6V should give $T_F=164\mu$K which is right at the peak g/b ratio in Figure-1. ii) we have a measurement at 0.35V showing a good to bad ratio of 67, scaling this to 0.6V also roughly gives g/b around 200. If we have g/b 200, why should the evaporation be so bad?)}
 For $p$-wave collisions, we have observed a new effect, the slow transfer of energy between low and high velocity groups.

An obvious question is whether Feshbach resonances can lead to a more favorable ratio of elastic to inelastic collisions.  Comprehensive studies of the $p$-wave resonance near 159\,G have been carried out by the Tokyo group\,\cite{MukaiyamaParameter,MukaiyamaScale}.  In the regime with universal scaling (Eq.\,4 in Ref.\,\cite{MukaiyamaScale}), the losses are 67 times higher than for background interactions.  However, effective range corrections can enhance elastic collisions on one side of the resonance and suppress them on the other side; they also increase losses.  An analysis of the results of Refs.\,\cite{MukaiyamaParameter,MukaiyamaScale} show a narrow range of magnetic field detunings where the ratio of good-to-bad collisions is around 10, but at rather slow elastic collision rate around 5/s (Appendix\,\ref{app:good to bad}).

An unexpected and intriguing observation is a quasi-equilibrium spin mixture of states $\ket{1}$ and $\ket{6}$.  Preliminary thermalization experiments show fast thermalization rates as high as $10^3$/s which could possibly enable effective evaporative cooling.  It may be possible to use state $\ket{6}$ as sensitive co-thermometer for a deeply degenerate cloud of state $\ket{1}$ atoms.
\begin{acknowledgments}
Acknowledgements: We thank Hyungmok Son for comments on the manuscript. We acknowledge support from the NSF through the Center for Ultracold Atoms and Grant No. 1506369, ARO- MURI Non-Equilibrium Many-Body Dynamics (Grant No. W911NF-14-1-0003), AFOSR- MURI Quantum Phases of Matter (Grant No. FA9550- 14-10035), ONR (Grant No. N00014-17-1-2253), and from a Vannevar-Bush Faculty Fellowship.
\end{acknowledgments}

 \appendix
 \section{Ratio of good to bad collisions}\label{app:good to bad}
 For the condition of our experiment, the zero-range approximation is valid, and the ratio of good to bad collisions for the experimentally determined collision parameters ($L_3$, $V_p$) is shown in  Fig.\,\ref{fig:8} of the main text. In the main text, we showed that the ratio of good to bad collisions scales as $1/(C V_p^{2/3} k^2)$.  If we rewrite the scaling as $(T_F /\Gamma_{el})^{1/3}/C$, and assume that a certain elastic collision rate is needed to overcome residual gas collisions, then the collision ratio depends only on the achievable densities (or $T_F$) and the constant $C$.  We have measured the constant $C$ for background scattering to be almost two orders of magnitude less than observed near a Feshbach resonance. Therefore, in the regime where the zero-range approximation applies, the $p$-wave Feshbach resonance near 159 G has an unfavorable ratio of good to bad collisions.
 
 However, near the Feshbach resonance, the zero-range approximation breaks down.  For elastic collisions, the effective range term interferes constructively with the $p$-wave scattering volume on the high-field side of the resonance (and destructively on the other side), which can improve the collision ratio, as we show here.

The thermally averaged elastic collision rate is given by
\begin{equation}
    \Gamma_{\text{el}} = n_{\text{mean}} \braket{\sigma_{pFR}(k) v_{\text{rel}}}
\end{equation}
where $v_{\text{rel}}$ is the relative velocity between colliding atoms, and $n_{\text{mean}}=\braket{n} = \frac{1}{N}\int n^2(\vec r) d^3r$ is the mean density (sometimes referred to as the density-weighted density), given by $n_{\text{mean}}=\frac{1}{48}(\frac{k_B m }{\hbar^2 \pi})^{3/2}\frac{T_F^{3}}{T^{3/2}}$ for an harmonically trapped Boltzman gas.  The $p$-wave cross section $\sigma_{pFR}(k)=24\pi\lvert f_{pFR}(k)\rvert^2$ is expressed by the scattering amplitude $f_{pFR}(k)$.  Near a $p$-wave Feshbach resonance, one can express the scattering amplitude using the effective range expansion\,\cite{UedaEffRange} as
\begin{equation}\label{eq:scattering amplitude}
    f_{pFR}(k) = \frac{k^2}{-\frac{1}{V(B)}+k_ek^2-ik^3}
\end{equation}
where $k_e$ is the second coefficient in the effective-range expansion, and the scattering volume varies as $V(B) = V_{p}(1+\frac{\Delta B}{B-B_{res}})$. We use the values from Ref. \cite{MukaiyamaParameter} to parametrize the Feshbach resonance: $V_{p}=(-41a_0)^3$, $k_e = -0.058 a_0^{-1}$ and $\Delta B = 40G$.

Inelastic collisions near the $p$-wave Feschbach resonance have been parametrized through the scaling relation
\begin{align}\begin{split}
    \Gamma_{\text{inel}} = \braket{n^2} L_3 = \braket{n^2} C \frac{\hbar}{m}k_T^4V_B^{8/3}
\end{split}\end{align}
where $\langle n^2\rangle = \frac{1}{N}\int n^3(\vec{r})d\vec{r}$ is the mean squared density, given by $\braket{n^2}=\frac{1}{864\sqrt{3}}(\frac{k_B m }{\hbar^2 \pi})^{3}\frac{T_F^6}{T^3}$ for a harmonically trapped Boltzman gas, $k_T=\sqrt{3mk_BT/2\hbar^2}$, $C=C_0\{1+(\beta k_e k_T^2 V_B)^\gamma\}$, $C_0 = 2\times 10^6$, $\beta=9$, $\gamma=14$\,\cite{MukaiyamaScale}.

With a vacuum limited lifetime of $\tau_{\text{vac}}=60$ seconds, we define ratio of good to bad collisions as

\begin{equation}
    \frac{\Gamma_{\text{el}}}{\Gamma_{\text{inel}}+1/\tau_{\text{vac}}}.
\end{equation}

\begin{figure}[b]
\includegraphics[width=0.5\textwidth]{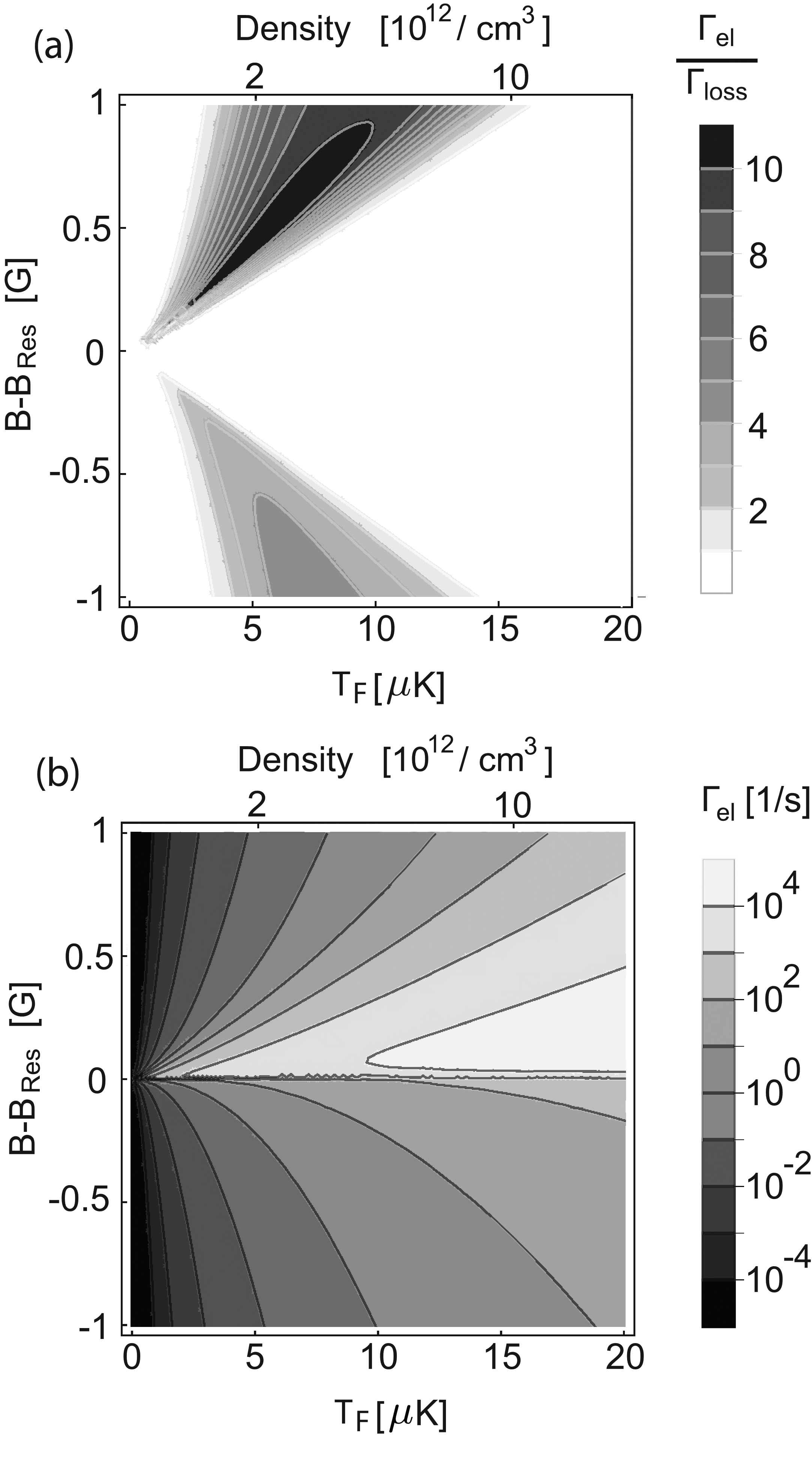}% Here is how to import EPS art
\caption{\label{fig:7} The ratio of good to bad collisions (Panel-(a)) and the elastic collision rate near the $p$-wave Feshbach resonance at 159 G  for a harmonically trapped cloud with $T=T_F$.  The ratio is worse below the resonance.}
\end{figure}

Figure\,\ref{fig:7}(a) shows the ratio of good to bad collisions around the Feshbach resonance in the range relevant to the Tokyo group experiment\,\cite{MukaiyamaParameter}. The maximum of the ratio is around 12. Figure\,\ref{fig:7}(b) shows the elastic collision rate.

\section{Analysis of three-body loss}\label{app:three body}

%=-L_3\frac{N^2}{24\sqrt{3}\pi^3\sigma_x^2\sigma_y^2\sigma_z^2} 
\emph{Harmonic approximation.} Since there are no inelastic two body collisions for $^{6}$Li atoms in the lowest hyperfine state (i.e. the two-body loss coefficient $L_2=0$), the time dependence of the number of atoms $N$ in the trap is described by the differential equation $\frac{\dot{N}}{N}=-L_3\langle n^2\rangle$. Later, we will add a single-particle loss rate $L_1$. For a non-degenerate cloud, and using the harmonic approximation for the trapping potential, the density profile is Gaussian, and we obtain
\begin{equation}\label{eq:harmonic loss}
\frac{\dot{N}}{N}=  -L_3\frac{N^2} {24\sqrt{3}\pi^3 \sigma_x^2\sigma_y^2\sigma_z^2} = -L_3\frac{N^2} {24\sqrt{3}\pi^3 V^2},
\end{equation}
where $\sigma_i = \sqrt{k_BT/m \omega_i^2}$ is the cloud's size in direction $i$, and the volume is defined as $V^2=\sigma_x^2\sigma_y^2\sigma_z^2$. The Wigner threshold law predicts that $L_3$ should scale with wavenumber $k$ as $L_3\sim k^4$\,\cite{PhysRevA.65.010705}. Since three-body losses also cause significant heating during the measurement, $L_3$ and the volume $V$ are both time-dependent, which has to be included in the analysis. Integrating both sides of the differential equation, we obtain:

\begin{align}\begin{split}
    \frac{\dot{N}}{N^3}&=-\frac{1}{24\sqrt{3}\pi^3}\frac{L_3(t)}{V^2(t)} \\
    \frac{1}{2N^2}&=\frac{1}{24\sqrt{3}\pi^3}\int_0^t\frac{L_3(t^\prime)}{V^2(t^\prime)}dt^\prime+\frac{1}{2N_0^2}
\end{split}\end{align}

We define $\Gamma_{3b}(t)=\frac{L_3(t)}{V^2(t)}$. Further simplifying the equation:

\begin{align}\begin{split}
    \frac{1}{2N^2}&=\frac{1}{24\sqrt{3}\pi^3}\Gamma_{3b}(0)\int_0^t\frac{\Gamma_{3b}(t^\prime)}{\Gamma_{3b}(0)}dt^\prime+\frac{1}{2N_0^2} \\
 \label{eq:ThreeBodyLoss}
    N(\tilde{t}) &=\sqrt{\frac{1}{\frac{1}{12\sqrt{3}\pi^3}\Gamma_{3b}(0)\tilde{t}+\frac{1}{N_0^2}}},
\end{split}\end{align}

where
\begin{equation}
\tilde{t}=\int_0^t\frac{\Gamma_{3b}(t^\prime)}{\Gamma_{3b}(0)}dt^\prime = \int_0^t \frac{T(0)}{T(t^\prime)} dt^\prime
\end{equation}

is the rescaled time, $T(t)$ the time-dependent temperature, and we have used $\Gamma_{3b}\sim 1/T$. Knowledge of the temperature evolution from the experimental measurements allows us to calculate $\tilde{t}$, which can be used in the new time axis in the fit.

%%%%%
\emph{Beyond the harmonic approximation.} The optical trapping potential is approximately Gaussian in the radial direction, and Lorentzian axially.  If the ratio of the trap depth to the temperature $\eta\equiv U_0/k_B T$ is smaller than $\sim$10, anharmonic corrections become important, and  the expression in Eq.\,\ref{eq:harmonic loss} overestimates the density. One then has to numerically calculate the mean squared density $\braket{n^2}$, as we discuss below.

%We assume a Boltzmann distribution for the population of quantum states with energy $E_i$, $\braket{n_i} \sim e^{-\beta E_i}$.
For temperatures $k_B T \gg \hbar \omega $, we can use a semiclassical description where the occupation of a phase space cell $\{ \vec{r},\vec{p} \}$ is given by a Boltzmann distribution: $f(\vec{r},\vec{p}) \propto \exp\left[-\beta\left(\vec{p}^2/2m + U(\vec{r})\right)\right]$,
where $U(\vec{r})$ is the external potential. The density distribution of the thermal gas is then given by integrating over the momentum:
\begin{align}\begin{split}
n_{th}(\vec r) &= \frac{1}{C}\int \frac{d^3\vec{p}}{(2\pi \hbar)^3}f(\vec{r},\vec{p}) \\&= \frac{1}{C}\frac{(m k_B T)^{3/2}}{2\sqrt{2}\pi^{3/2}\hbar^3}\exp[-\beta U(\vec{r})].
\end{split}\end{align}
The normalization constant $C$ is determined by the total number of atoms $N$:
\begin{equation}
C = \frac{1}{N} \int \frac{(m k_B T)^{3/2}}{2\sqrt{2}\pi^{3/2}\hbar^3}\exp[-\beta U(\vec{r})] d^3\vec{r}.
\end{equation}
%Substituting $C$ into the expression for $n_{th}$, we obtain
%\begin{equation}
%n_{th} = \frac{N\exp[-\beta U(\vec{r})]}{\int \exp[-\beta U(\vec{r})] d^3\vec{r}}.
%\end{equation}
 %The density-weighted density is then given by
%\begin{align}\begin{split}
%\braket{n} &= \frac{1}{N}\int n_{th}(\vec{r})^2 d^3r 
%= \frac{1}{N} \frac{N^2\int \left (\exp[-\beta V(\vec{r})] \right)^2 d^3r}{\left( \int \exp[-\beta V(\vec{r})] d^3r %\right)^2}\\
%&=N \frac{\int \exp[-2\beta V(\vec{r})] d^3r}{\left( \int \exp[-\beta V(\vec{r})] d^3r \right)^2} = \frac{N}{2\pi} %\frac{\iint r\exp[-2\beta V(\vec{r})] dr dz}{\left( \iint r\exp[-\beta V(\vec{r})] dr dz \right)^2}.
%\end{split}\end{align}

The mean squared density $\langle n^2\rangle$ is then given by
\begin{align}\begin{split}
\braket{n^2} &= \frac{1}{N}\int n_{th}(\vec{r})^3 d^3r \\
&= \frac{N^2}{(2\pi)^2} \frac{\iint r\exp[-3\beta U(\vec{r})] dr dz}{\left( \iint r\exp[-\beta U(\vec{r})] dr dz \right)^3},
\end{split}\end{align}
where $rdrdz$ is the volume element in cylindrical coordinates.
%&=N^2 \frac{\int \exp[-3\beta U(\vec{r})] d^3r}{\left( \int \exp[-\beta U(\vec{r})] d^3r \right)^3} \\
%\textcolor{red}{some intermediate equations can be eliminated}

Without the harmonic approximation of $U(\vec r)$ this integral is generally not analytically solvable. We evaluate it numerically, using the explicit potential of a focused Gaussian beam \cite{VarennaNotes}:
\begin{align}\begin{split}
&U_{\text{ODT}}(\vec{r}) = -\frac{U_0}{1+(z/z_R)^2}\exp\left( -\frac{2r^2}{w(z)^2} \right),
\end{split}\end{align}
%&= -\frac{3 \pi c^2}{2 \omega_0^3} \left(\frac{\Gamma }{\omega_0-\omega}+\frac{\Gamma }{\omega_0 +\omega}\right)\frac{2P}{\pi w(z)^2}\exp\left( -\frac{2r^2}{w(z)^2} \right)\\
where $w(z) = w_0\sqrt{1+(z/z_R)^2}$ is the beam spot size, $z_R$ is the Rayleigh range of the beam, and $U_0$ is the trap depth. %is given by $U_0 = |U_{\text{ODT}}(r=0, z=0)| = \frac{3 \pi c^2}{2 \omega_0^3} \left(\frac{\Gamma }{\omega_0-\omega}+\frac{\Gamma }{\omega_0 +\omega}\right)\frac{2P}{\pi w_0^2}$.
%Here $c$ is the speed of light, $\omega$ is the laser frequency, $\omega_0$ is the frequency of the atomic transition, $\Gamma$ is the natural linewidth of the transition, and $P$ is the laser power.

Since gravity tilts the potential, it sets a constraint on the maximum energy an atom can have inside the trap. We include this effect by setting the integration limits to 95\% of the full trap depth, i.e. $U_{\text{ODT}}(r_{max})=0.95U_0$, and $U_{\text{ODT}}(z_{max})=0.95U_0$. Without this cutoff, $\braket{n^2}$ tends towards zero.

To account for the scaling of the three-body loss coefficient with temperature, we modify Eq.\,\ref{eq:harmonic loss} such that $\dot{N} = L_1 N -L_3^T T^2 \braket{n^2} N$, where $L_3^T$ is the temperature-independent part of $L_3$. We insert the numerically calculated time-dependent $\braket{n^2}$ into the modified form of Eq.\,\ref{eq:loss}, and fit the decay of the atom number as a function of hold time. The values shown in Fig.\,\ref{fig:scattering length} represent the loss coefficient at the beginning of each measurement (i.e. at zero hold time).

\section{Calculation of the $p$-wave scattering rate}\label{app:therm calc}
The thermally averaged $p$-wave elastic collision rate is given by
\begin{align}\begin{split}
    \Gamma_{\text{el}}&=n_{\text{mean}}\langle \sigma_p(k) v_{\text{rel}} \rangle \\
    &=n_{\text{mean}}\sqrt{\frac{8}{\mu\pi}}(k_BT)^{-3/2}24\pi\int| f_p(k)|^2Ee^{-\beta E}dE.
\end{split}\end{align}
Far-away from the Feshbach resonance, and in the low energy limit, the scattering $p$-wave scattering amplitude (Eq.\,\ref{eq:scattering amplitude}) becomes $f_{pFR}(k) \approx -V_p k^2$. The $p$-wave scattering cross-section for identical fermions is then given by
\begin{equation}
\sigma_p(k) = 24\pi|f_p(k)|^2 = 24\pi V^2_p k^4,
\end{equation}
where $V_p$ is the $p$-wave background scattering volume. The integral over the energy then gives $V^2_p \frac{4\mu^2}{\hbar^4} \frac{24}{\beta^5}$,
%\begin{align}\begin{split}
%    \int| f_p(k)|^2Ee^{-\beta E}dE&=  \int V^2_p \frac{4\mu^2}{\hbar^4} E^3e^{-\beta E}dE \\
%    &= V^2_p \frac{4\mu^2}{\hbar^4} \frac{24}{\beta^5},
%\end{split}\end{align}
where we used $E=\hbar^2 k^2 /2\mu$ for the collision energy, and where $\mu$ is the reduced mass. Combining all terms, and using $\Gamma_{\text{th}}=\Gamma_{\text{el}}/\alpha$, one directly obtains Eq.\,\ref{eq:GammaTh}.
%\begin{align}
%    \Gamma_{\text{el}}&=1152\sqrt{2\pi}\frac{V^2_p \mu^{3/2}}{\hbar^4}n_{\text{mean}}(k_B T)^{5/2}
%\end{align} 
%\begin{align}\begin{split}
%    \Gamma_{\text{th}}&=\frac{1}{\alpha}\Gamma_{\text{el}} \\
%    &=\frac{1152}{\alpha}\sqrt{2\pi}\frac{V^2_p \mu^{3/2}}{\hbar^4}n_{\text{mean}}(k_B T)^{5/2}.
%\end{split}\end{align}

\section{Comparing $s$ and $p$-wave collision rates}\label{app:s and p comp}
\begin{figure}[b]
\includegraphics[width=0.5\textwidth]{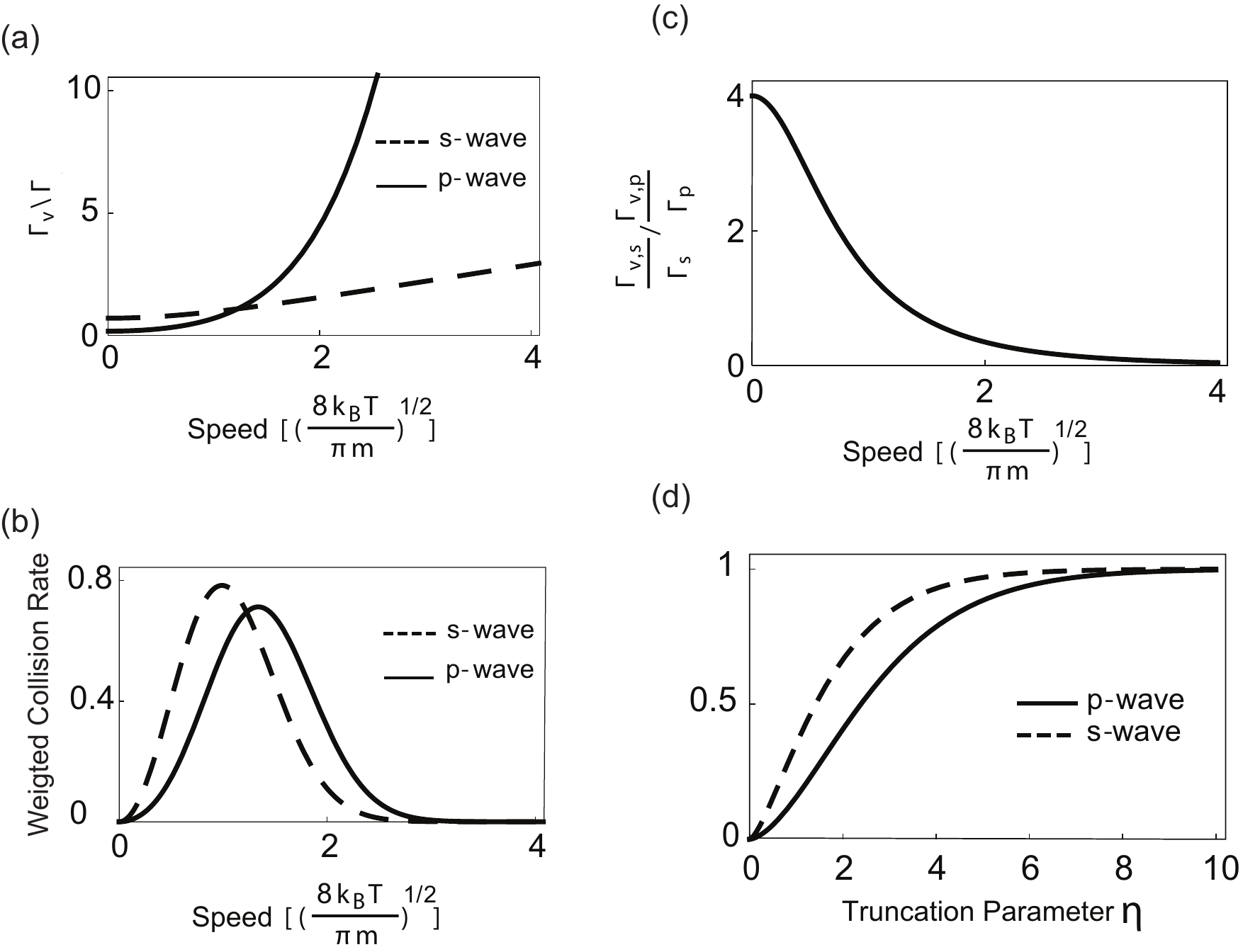}% Here is how to import EPS art
\caption{\label{fig:6} Comparison between $s$ and $p$-wave collision rates.
Panel (a) shows the rate of collisions involving an atom with speed $v$ with respect to the thermally averaged collision rate. In panel (b), the collision rates are weighted by the Maxwell-Boltzman speed distribution.
Panel (c) shows the ratio of $s$ and $p$ wave collisions for an atom at speed $v$.  Panel (d) is the fraction of collisions up to a kinetic energy of $\eta k_B T$ where $\eta$ is the truncation parameter.}
\end{figure}

Here we analyze how the $k^4$ dependence of the $p$-wave scattering cross-section
affects thermal equilibration, and compare to the familiar case of $s$-wave scattering.  The collision rate for partial wave $l$ involving an atom with speed $v$ is
%\begin{align}\begin{split}
%     \Gamma_{l,\lvert \vec{v_1}\rvert=v} = n_{\text{mean}} \int\int \sigma_l \lvert %\vec{v_1}-\vec{v_2} \rvert f_{v1}f_{v2}d^3\vec{v_1}d^3\vec{v_2}, \\
%\end{split}\end{align}
\begin{equation}
     \Gamma_{l,v} = n_{\text{mean}} \int \sigma_l \lvert \vec{v}-\vec{v_2} \rvert f_{v2} d^3v_2.\\
\end{equation}
where $\sigma_l$ is the collision cross-section, and $f_v$ is the Maxwell-Boltzmann velocity distribution.  %We are forcing $v_1$ to take a particular value. 
%properly normalized velocity distributions are
%\begin{align}\begin{split}
%    f_{v1} &= e^{\beta\frac{m}{2}v^2}\delta(v_{1x}-v_x)\delta(v_{1y}-v_y)\delta(v_{1z}-v_z)e^{-\beta\frac{m}{2}(v^2_{1x}+v^2_{1y}+v^2_{1z})} \\
%    f_{v2} &= (\frac{\beta m }{2\pi})^{3/2} e^{-\beta\frac{m}{2}(v^2_{2x}+v^2_{2y}+v^2_{2z})}
%\end{split}\end{align}
%Integrating over $v_1$ gives, 
We are interested in the ratio of the elastic scattering rate for an atom with velocity $v$ compared to the average scattering rate:
\begin{align}
    \frac{ \Gamma_{p,\lvert \vec{v_1}\rvert=v}}{\Gamma_p}&=\frac{\sqrt{2}}{48} \int^{\infty}_0\int^\pi_0(\tilde{u}^2+u^2-2\tilde{u}u\cos{\theta})^{5/2}\sin{\theta}d\theta e^{-u^2}u^2du, \\
    \intertext{and}
    \frac{ \Gamma_{s,\lvert \vec{v_1}\rvert=v}}{\Gamma_s}&=\frac{1}{\sqrt{2}}\int^{\infty}_0\int^\pi_0\sqrt{\tilde{u}^2+u^2-2\tilde{u}u\cos{\theta}}\sin{\theta}d\theta e^{-u^2}u^2du,
\end{align}
where $\beta\frac{m}{2}v^2=\tilde{u}^2$. Figure\,\ref{fig:6} shows these collision rates and their ratio.  Considering the mean speed is $\sqrt{\frac{8 k_B T}{\pi m}}$, $s$-wave interactions have three times more collisions for speeds below the average speed (Fig.\,\ref{fig:6}a). Detailed balance implies that the rate of collisions generating low-speed particles is the same as the rate of collisions involving a low-speed particle.  In evaporative cooling the high-energy tail of the atomic distribution is removed, and collisions repopulate the tail and transfer population to lower velocities.  For the same average collision rate, the transfer of population to the velocity group near $v=0$ takes 4 times longer for $p$ wave collisions compared to $s$ wave (panel (c)). In addition, for a trap depth of $\eta k_B T$ with $\eta=5$, the rate of $p$-wave collisions calculated for a full Boltzmann distribution overestimates the collision rate  of the truncated Boltzmann distribution by  13\% (panel (d)).  As a result, efficient evaporation with $p$-wave collisions requires a substantially higher ratio of good to bad collisions than $s$-wave collisions.

%\subsection{Dipolar relaxation of state $\ket{F=3/2,+3/2}$}
%We started our measurements with the spin polarized stretched $\ket{F=3/2,+3/2}$ state of $^6$Li. A careful examination of the sample showed dipolar relaxation of the the stretched state into the lower hyperfine state. An exapmple impurity generation curve is given in Figure-\ref{fig:4}. The speed of impurity generation and peak number of generated impurities are density dependent.

%PUT DETAILS WHICH DON"T FIT INTO THE MAIN TEXT

%\begin{figure}[b]
%\includegraphics[width=0.5\textwidth]{EvapSecond2.eps}% Here is how to import EPS art
%\caption{\label{fig:5} $p$-wave evaporation. Panel (a) is average of 4 shots for a sample with $T/T_F=0.52$. Panel (b) displays the evolution of $T/T_F$ vs the atom number. Panel (c) and (d) are gaussian fit to $1/e$ radius wings of the atomic cloud. The gaussian fit clearly over estimates the peak, a sign of lack of low velocity $p$-wave collisions and quantum degeneracy.} 
%\end{figure}

% The \nocite command causes all entries in a bibliography to be printed out
% whether or not they are actually referenced in the text. This is appropriate
% for the sample file to show the different styles of references, but authors
% most likely will not want to use it.
%\nocite{*}

\bibliography{apssamp}% Produces the bibliography via BibTeX.

\begin{thebibliography}{38}
\bibitem [{foo()}]{footnote}%
  \BibitemOpen
  \href@noop {} {}\bibinfo {note} {Ref. \cite{MukaiyamaParameter} mentions an
  experimental value for $V_p$. It was obtained by using the experimentally
  determined strength of a p-wave resonance which is 1.65 higher than
  calculated, and then multiplying a predicted value of $V_p$ by this factor.
  There is no obvious reason for rescaling the theoretical value for
  $V_p$.}\BibitemShut {Stop}%
\end{thebibliography}

\end{document}